\documentclass[12pt,letterpaper,copyedit]{article}
\usepackage{osajnl}
\usepackage{graphicx}
\usepackage{bm}
\usepackage{amsmath,amsfonts,epsfig,amssymb}
\usepackage{cite}

\begin{document}

\title{A Radio-Frequency Atom Chip for Guiding Neutral Atoms}

\author{Xiaolin Li, Haichao Zhang, Bo Yan, Min Ke, Yuzhu Wang}

\address{Key Laboratory for Quantum Optics, Shanghai Institute
of Optics and Fine Mechanics, Chinese Academy of Sciences, Shanghai,
China, 201800}

\email{opticalman@gmail.com}

\maketitle

\begin{abstract}\label{txt:abstract}
We propose two kinds of wire configurations fabricated on an atom
chip surface for creating two-dimensional (2D) adiabatic rf guide
with an inhomogeneous rf magnetic field and a homogenous dc magnetic
field. The guiding state can be selected by changing the detuning
between the frequency of rf magnetic field and the resonance
frequency of two Zeeman sublevels. We also discuss the optimization
of loading efficiency and the trap depth and how to decide proper
construction when designing an rf atom chip.
\end{abstract}


\section{Introduction}
\label{intro}

Magnetic field is widespread used for manipulating neutral atoms and
becomes a powerful tool for studying the physics of atoms
\cite{Wieman99}. Accurate metal wires at a scale of micron can be
fabricated in a chip with lithographic or other standard
surface-patterning process \cite{Drndi98}. This kind of chip can
realize miniaturized electric or magnetic traps and guides which can
manipulate neutral atoms close to the chip surface
\cite{Reichel01,Ott01,Folman00,Muller99,Dekker00,Du04}. Compared to
the common magnetic coil, the atom chip can not only create tightly
atom traps with large field gradients and high field curvature
without the need for large current and large volume, but versatile
and tiny traps without complex structure and design. This permits
the controlled tunneling of atoms over micron or sub-micron lengths
and makes it a natural platform for applications in coherent
matter-wave control such as miniaturized atom interferometer,
quantum information process, and the study of low-dimensional
quantum gases \cite{Fortagh00}. But ordinary atom chips use static
magnetic field for atom trapping, only weak-field-seeking atoms
(excited spin state atom) can be trapped since strong-field seeking
atoms (ground spin state atoms) need the maxima of magnetic field
and this kind of magnetic field is not allowed by Maxwell's
equations. For weak-field-seeking atoms, spin of excited spin states
relax to the ground state at a density-dependent rate, afterwards
these atoms are ejected from the trap \cite{Lagendijk86}. This can
be overcame by using ac magnetic field. R. Lovelace et al
\cite{Lovelace85} proposed a dynamic magnetic trap using ac magnetic
field for trapping which can trap strong-field seeking atoms. Then
C. Agosta et al \cite{Agosta89} proposed another scheme using
microwave radiation which is realized experimentally by R. Spreeuw
et al \cite{Spreeuw94}, but the equipment is complex and difficult
to realize. In comparison with microwave, rf technique has been used
widespread in forced evaporative cooling and other applications of
neutral atoms and easily manipulated in experiment. Recently the rf
magnetic field is introduced to realize rf adiabatic trap combining
with a dc Ioffe-Pritchard trap\cite{Zobay01,Lesanovsky06,Schumm05}.
In comparison to the optical trap, spontaneous radiation can be
ignored in this kind of rf magnetic trap and the energy level is
finite and can be calculated easily.

In this article we propose an rf atom chip for guiding atoms in
strong-field seeking state. In Section \ref{sec:1} we discuss the
theory of adiabatic rf guide and then two kinds of guide
configurations are proposed. In Section \ref{sec:2} parameters of
the guides and loading efficiency are also discussed. In Section
\ref{sec:3} we discuss the problems and solutions concerning design
of an rf atom chip.

\section{Theory of Adiabatic Radio-Frequency Guide}
\label{sec:1}

Now the dressed state method is introduced for calculating the
eigenstate and eigenenergy. We consider a neutral $^{87}Rb$ atom in
ground state $|F=2\rangle$ which has five zeeman sublevels
($m_{F}=-2,-1,0,1,2$) . Hamiltonian of an atom in rf magnetic field
with a dc bias magnetic field is given by

\begin{equation}
\hat{H}=\hat{H}_{radio}+\hat{H}_{atom}+V(\mathbf{r}),\label{eqn:ham}
\end{equation}

where $\hat{H}_{radio}$ is the Hamiltonian of the rf field,
$V(\mathbf{r})$ is the atom-field coupling strength which is given
in magnetic-dipole and rotating-wave approximation by
$V(\mathbf{r})=\frac{\lambda}{\sqrt{2}}(aS_{+}+a^{+}S_{-})$ with
$S_{\pm}=S_{x}\pm i S_{y}$. $\hat{H}_{atom}$ is the atomic
Hamiltonian which is given by

\begin{equation}
\hat{H}_{atom}=\frac{\mathbf{P}^{2}}{2M}-\pmb{\mu}
\cdot\mathbf{B}_{0} \equiv \frac{\mathbf{P}^{2}}{2M}+\hbar \omega_0
S_{y}, \label{eqn:hamatom}
\end{equation}

where $\pmb{\mu}$ is the magnetic dipole moment of the atom,
$\omega_0=g_F \mu_B B_0$ is the resonance frequency with the Bohr
magneton $\mu_B$, The Land\'{e} factor $g_{F}$ and the homogeneous
dc magnetic field $\mathbf{B}_0=B_0 \mathbf{e}_y$ along the Y-axis
as shown in Fig.\,\ref{fig:section}. And $S_y$ is the y component of
$\mathbf{S}$ which is given by atomic angular momentum
$\mathbf{F}=\hbar \mathbf{S}$. Supposing that atomic velocity is so
small that atom motion cannot excite the transitions among the
internal states, one can use Born-Oppenheimer approximation to
separate the whole wave-function $|\psi \rangle$ of the global
system into two parts: one is the external state
$\phi_n(\mathbf{r})$ of atom motion and another is the internal
state $\chi_n(\mathbf{r})$ of atom in rf field. Afterwards, two
separated but correlative equation is given by
\begin{eqnarray}
H_{DA} ({\mathbf{r}})\left| {\chi _n ({\mathbf{r}})} \right\rangle  = \varepsilon _n ({\mathbf{r}})\left| {\chi _n ({\mathbf{r}})} \right\rangle  \hfill ,\\
  (\frac{{{\mathbf{P}}^2 }}
{{2M}} + \varepsilon _n ({\mathbf{r}}))\phi _n ({\mathbf{r}}) =
E\phi _n ({\mathbf{r}}) \hfill,
\end{eqnarray}
where $H_{DA} ({\mathbf{r}}) \equiv H - \frac{{{\mathbf{P}}^2 }}
{{2M}} = \hat H_{Radio}  + \hbar \omega _0 S_y  + V({\mathbf{r}})$
is dressed atom Hamiltonian with its eigenvalues of $\varepsilon _n
({\mathbf{r}})$. All of five dressed eigenenergies of
$H_{DA}(\mathbf{r})$ can be obtained as
\begin{equation}
\varepsilon_{m_{F}}=(N+2)\hbar \omega +
m_{F}\hbar\sqrt{\Omega^{2}(\mathbf{r})+\delta^{2}},\label{eqn:eignpo}
\end{equation}
where $m_F=2,1,0,-1,-2$, $\Omega(\mathbf{r})=\mu_{B}B_{rf}/(2\hbar)$
and the detuning $\delta=\omega-\omega_0=\omega-g_F \mu_B
B_0/\hbar$. It should be paid attention to that the result uses the
rotating wave approximation. The proposal of O.~Zobay et al
\cite{Zobay01} starts from a static Ioffe-Pritchard trap and then
superimposes a homogeneous oscillatory rf field where
$\Omega(\mathbf{r})$ is homogeneous and $\delta$ has inhomogeneous
spatial distribution. Our proposal reverse the idea and starts from
a homogenous magnetic field (bias field $B_0$) and superimposes an
oscillatory rf field varying in the space. The results will go
beyond the reversion and there will come some new phenomena. As
shown in Equation (\ref{eqn:eignpo}), for the rf magnetic field with
a minimum at its center (simplest configuration is a quadrupole rf
magnetic field), only $\varepsilon_1(\mathbf{r})$ and
$\varepsilon_2(\mathbf{r})$ have a minimum at minimum point of rf
field, so their eigenstate become trapped state which is given by
\begin{eqnarray}
 \left| {\chi _1 } \right\rangle  &&= \cos ^4 \theta \left| {\varphi _2 } \right\rangle  + 2\cos ^3 \theta \sin \theta \left| {\varphi _1 }
 \right\rangle \nonumber \\
 &&+ \sqrt 6 \cos ^2 \theta \sin ^2 \theta \left| {\varphi _0 } \right\rangle   \nonumber \\
 &&+2\cos \theta \sin ^3 \theta \left| {\varphi _{ - 1} } \right\rangle  + \sin ^4 \theta \left| {\varphi _{ - 2} }
 \right\rangle,
 \label{eqn:chi1}\\
 \left| {\chi _2 } \right\rangle  &&= 2\cos ^3 \theta \sin \theta \left| {\varphi _2 } \right\rangle    - \frac{1}{2}(\cos 2\theta  + \cos 4\theta )\left| {\varphi _1 } \right\rangle \nonumber\\
 &&- \sqrt {\frac{3}{8}} \sin 4\theta \left| {\varphi _0 } \right\rangle +\frac{1}{2}(\cos 4\theta  - \cos 2\theta )\left| {\varphi _{ - 1} }
\right\rangle \nonumber \\
&& - 2\cos \theta \sin ^3 \theta \left|
{\varphi _{ - 2} }
 \right\rangle ,\label{eqn:chi2}
\end{eqnarray}
where the angle
\begin{equation}
\theta=\frac{1}{2}arccos(\frac{1}{\sqrt{(\Omega(\mathbf{r})/\delta)^{2}+1}})
\label{eqn:theta}
\end{equation}
and $|\varphi_{m_{F}}\rangle\equiv|m_F,N+F-m_F\rangle$  is
eigenstate of $H_{DA}(\mathbf{r})$ when atom-field coupling
$V(\mathbf{r})=0$ with $F=2$,$m_F=2,1,0,-1,-2$, and the photon
occupation number N. When $\delta/\Omega\ll1$,
$\theta\rightarrow\pi/2$ and $\left| {\chi _1 } \right\rangle$ and
$\left| {\chi _{2} } \right\rangle$ approach $\left| {\varphi _{-2}
} \right\rangle$ and $\left| {\varphi _{-1} } \right\rangle$
respectively . Then the trapped state is predominated by the
$|F=2,m_F=-2\rangle$-like state. If we prepare atoms in
$|F=2,m_F=-2\rangle$ Zeeman sublevel by optical pumping, atoms will
project into trapped state $\left| {\chi _1 } \right\rangle$ mostly,
the less part into $\left| {\chi _2 } \right\rangle$ and little
residual into untrapped states after they couple with a quadrupole
rf magnetic field. This differs from the static magnetic trap that
can only trap the states with $m_{F}>0$ and $g_{F}>0$ or $m_{F}<0$
and $g_{F}<0$. Actually $\delta/\Omega\ll1$ is not always true which
will make more atoms projecting into untrapped states and atom loss
becomes notable. We will study this problem in Section \ref{sec:2}.
And it must be pointed out that only the rf field components being
perpendicular to the dc bias magnetic field $B_0$ contribute to the
trap which means only a 2D guide will be created and there is no
trapping potential along the direction of $B_0$ \cite{Lesanovsky06}.

\section{Adiabatic Radio-Frequency Guide from Planar Wires}
\label{sec:2}

\begin{figure}
\centering
\includegraphics[width=8cm]{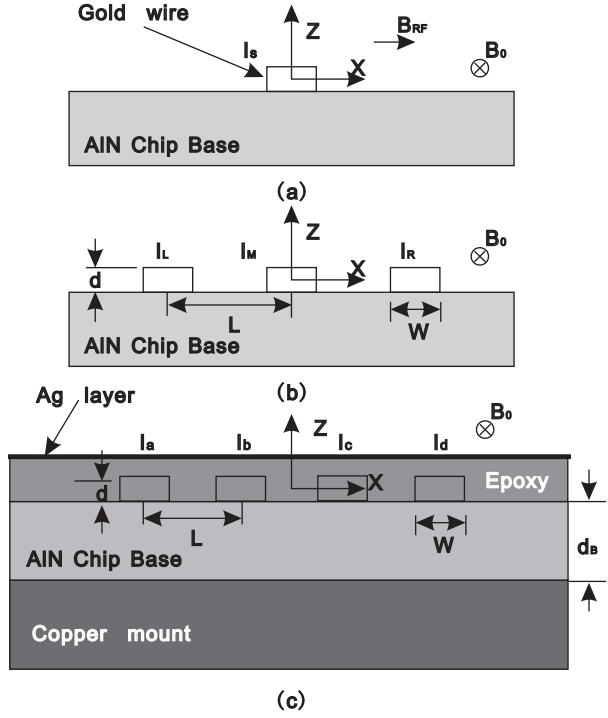}
\caption{Section figure of the atom chip. Gold wires with
rectangular section (wire width w and height d) are fabricated on a
chip and one-wire configuration is showed in (a) where the wire on
the chip carries Rf current ($I_{s}=I_{0}sin(\omega t)$). In
three-wire configuration (b), $I_{L}=I_{R}=I_{0}sin(\omega t),
I_{M}=I_{0}sin(\omega t+\pi)$. In four-wire configuration
(c),$I_{a}=I_{c}=I_{0}sin(\omega t), I_{b}=I_{d}=I_{0}sin(\omega
t+\pi)$} \label{fig:section}
\end{figure}

As shown above, in addition to a homogenous static magnetic field
($B_{0}$ in Fig.\,\ref{fig:section}), a 2D Radio-Frequency
quadrupole field is needed for the rf guide which is usually created
by a pair of anti-Helmholtz coils. But there will be a contradiction
between the huge inductive reactance in radio-frequency range and
the adequate current for sufficient trap depth and frequency. It is
difficult to get impedance match between huge inductive reactance
coils and a rf amplifier since the coils play a role of low-pass
filter. Atom chip provides with a new solution for the dilemma which
is made of planar wires fabricated in a microelectronic chip.
Because atom chip consists of several metal wires on the surface of
a chip, there will be little inductive and capacitive impedance for
impedance matching easily, as well as a little resistance for small
dissipation power. And the atom chip can provide huge gradient and
curvature with relatively small current since they respectively
scale as $I/s^2$ and $I/s^3$ ($I$ is current and $s$ is
characteristic size) \cite{Reichel02}. Although the conclusion is
made in direct current range, for the rf regime, if the distance
from the chip $r \ll \lambda$ ($\lambda$ is the wavelength of rf
magnetic field), the problem will transform to a static field
problem and the field amplitude can be calculated as a static
magnetic field \cite{Guo97}. Now the frequency of the rf magnetic
field is $\sim$ 10MHz and corresponding to a wavelength of $\sim$
30m which meets the condition.

The simplest configuration for the 2D quadrupole field in an atom
chip is a wire and a bias rf magnetic field perpendicular to the
wire (Fig.\,\ref{fig:section}a), which has been used for dc
guided\cite{Dekker00,Thywissen99,Cassettari00,Gunther05}. There will
be a zero-field point above the wire and a 2D quadrupole field
around it in XZ-plane. The distance between the chip and the
zero-field point depends on the ratio of the current and the
strength of the bias field \cite{Reichel02}. But the bias rf
magnetic field is created by the outer coils which will encounter
the same problem mentioned above. The solution is using wires in
atom chip for creating the bias rf magnetic field instead of outer
coils, as shown in Fig.\,\ref{fig:section}b and c, both three-wire
and four-wire configuration are able to create requisite 2D rf
quadrupole field.

\begin{figure}
\centering
\includegraphics[width=8cm]{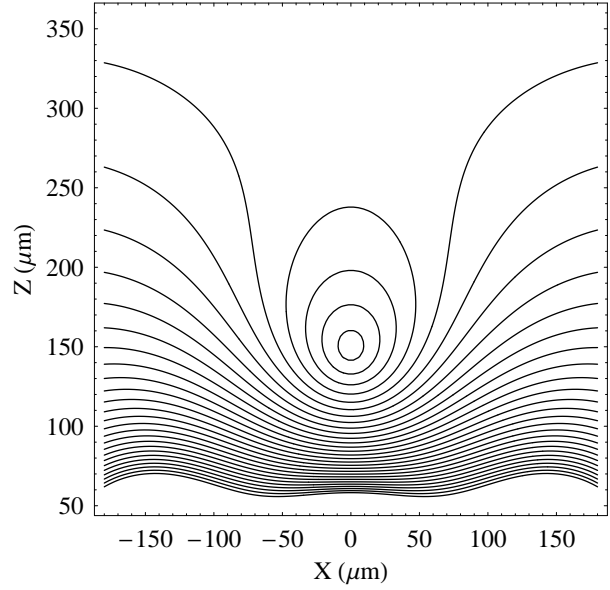}
\caption{Contour figure of adiabatic rf potential created by a
3-wire configuration where separation between two wires $l=150\mu
m$, current amplitude $I_{0}=2$A and current frequency detuing
$\delta=1$MHz.} \label{fig:3wp}
\end{figure}

Firstly wires are considered as infinitely thin and long, which is a
good approximation when characteristic distance from the chip
surface is bigger than the wire width and great smaller than the
wire length \cite{Reichel02}. Rf current is carried by the wires
where are shown in Fig.\,\ref{fig:section}b and c and dc field
$B_{0}$ is created by outer coils. According to equation
(\ref{eqn:eignpo}), in the case of trapped state $|\chi_{1}\rangle$,
the adiabatic rf potential of three-wire and four-wire configuration
respectively,

\begin{equation}
U_{{\rm{3 - wire,4 - wire}}}  = \mu _B \sqrt {(\frac{{\hbar\delta
}}{{\mu _B }})^2  + \frac{1}{4}(B_{3 - wire,4 - wire} )^2 },
\label{eqn:u34}
\end{equation}

and rf magnetic field $B_{3-wire}$ and $B_{4-wire}$ are given by

\begin{eqnarray}
 B_{3 - wire} ^2  &&= \frac{1}{4}(\frac{{\mu _0 I_0 }}{{2\pi }})^2 \{ [\sum\limits_{n =  - 1}^1 {( - 1)^{n + 1} \frac{z}{{(x + nl)^2  + z^2 }}} ]^2  \nonumber\\
  &&+ [\sum\limits_{n =  - 1}^1 {( - 1)^{n + 1} \frac{{x + nl}}{{(x + nl)^2  + z^2 }}} ]^2 \},\label{eqn:b3}\\
 B_{4 - wire} ^2  &&= \frac{1}{4}(\frac{{\mu _0 I_0 }}{{2\pi }})^2 \{ \{ \sum\limits_{n =  - 1}^2 {( - 1)^n \frac{z}{{[x + (n - 0.5)l]^2  + z^2 }}} \} ^2  \nonumber\\
  &&+ \{ \sum\limits_{n =  - 1}^2 {( - 1)^n \frac{{x - (n - 0.5)l}}{{[x + (n - 0.5)l]^2  + z^2 }}} \} ^2 \}.\label{eqn:b4}
 \end{eqnarray}
Fig.\,\ref{fig:3wp} shows adiabatic rf potential in XZ-plane.

Afterwards the guide centers of three-wire and four-wire
configuration can be found at $(0,l)$ and $(0,\frac{\sqrt{3}}{2}l)$
respectively, where both $U_{3-wire}$ and $U_{4-wire}$ are in the
minimum point. The guide centers are independent with current
amplitude and detuning $\delta$ and merely decided by the geometry
configuration. Around the guide centers, Taylor series of adiabatic
rf potential $U_{3-wire}, U_{4-wire}$ is respectively given by
($\delta\neq0$)

\begin{eqnarray}
U_{3 - wire} (z - l) &&= \mu _B b + \frac{{I_0 ^2 \mu _0 ^2 \mu _B
}}{{32l^4 \pi ^2 b}}(z - l)^2 \nonumber\\
&&+ O[z - l]^3,   \label{eqn:u3z}\\
 U_{3 - wire} (x) &&= \mu _B b + \frac{{I_0 ^2 \mu _0 ^2 \mu _B }}{{32l^4 \pi ^2 b}}x^2   + O[x]^3,  \label{eqn:u3x} \\
 U_{4 - wire} (z - \frac{{\sqrt 3 }}{2}l) &&= \mu _B b + \frac{{I_0 ^2 \mu _0 ^2 \mu _B }}{{24l^4 \pi ^2 b}}(z -
 l)^2 \nonumber\\
 &&+ O[z - l]^3,  \label{eqn:u4z} \\
 U_{4 - wire} (x) &&=  \mu _B b + \frac{{I_0 ^2 \mu _0 ^2 \mu _B }}{{24l^4 \pi ^2 b}}x^2  + O[x]^3. \label{eqn:u4x}
 \end{eqnarray}

$b$ is given by
\begin{equation}
b = \hbar\delta /\mu _B.
\end{equation}

According to the series, there appears constant and second terms and
no linear item, thereby the adiabatic rf potential $U_{3-wire}$ and
$U_{4-wire}$  approximates harmonic potential which is widespread
available in static magnetic potential for cold atoms and BEC. The
oscillation frequency along $i$th coordinate axis of $U_{3-wire}$
and $U_{4-wire}$ is given by (m is atom mass)
\begin{eqnarray}
\omega _{3 - wire;x,z}  \approx \frac{{\sqrt 3 }}{2}\omega _{4 -
wire;x,z} \nonumber \\
\approx \frac{{\sqrt {2hm} \mu _0 \mu _B }}{{8\pi hm}}\frac{{I_0
}}{{l^2 \sqrt \delta  }} \approx 0.67
(m^2s^{-1/2}A^{-1})\times\frac{{I_0 }} {{l^2 \sqrt \delta
}}(Hz).\label{eqn:tfreq}
\end{eqnarray}
It is apparent that the oscillation frequency of the three-wire and
four-wire configuration has a fixed ratio of $\sqrt{3}/2$ and each
of them has the same oscillation frequency along $X$ and $Z$ axis.
We substitute the atom mass of Rubidium for $m$ and will derive a
coefficient of $0.67$ (as shown in Equation (\ref{eqn:tfreq})). The
oscillation frequency is proportional to current amplitude $I_{0}$
and inversely proportion to square of $l$ and the root of $\delta$.
When $\delta$ becomes very small, the trap potential approach to a
linear and the guide approaches to a 2D quadrupole guide and a huge
oscillation frequency will be obtained. But the parameters can not
be decided optionally and analysis as follows will disclosure the
problem. After analysis of trap depth and transfer efficiency from
$|F=2,m_{F}=-2\rangle$ Zeeman state to all of the trapped dressed
states, a contradiction will be found which will make us select
detuning $\delta$ carefully.

According to first derivative of Equation \,(\ref{eqn:u34}),we can
find the maximum and minimum point of adiabatic rf potential and
then trap depth is decided easily. It is interesting that the trap
depth has the same form except a coefficient whether in three-wire
and four-wire configuration or along X and Z axis. Afterwards the
trap depth can be written in one equation which is given by
\begin{equation}
U_{i,j}  = \sqrt {k_{i,j} I_0 ^2 \mu _0 ^2 \mu _B ^2 /l^2  + \hbar^2
\delta ^2 }  - \hbar\delta. \label{eqn:utd}
\end{equation}
$k_{i,j}$ is the coefficient which is different in various situation
and is given by
\begin{eqnarray}
  k_{3 - wire,X}  = 3.34 \times 10^{ - 3}  \hfill,\nonumber\\
  k_{4 - wire,X}  = 5.34 \times 10^{ - 3}  \hfill, \nonumber\\
  k_{3 - wire,Z}  = 5.71 \times 10^{ - 4}  \hfill, \nonumber\\
  k_{4 - wire,Z}  = 4.55 \times 10^{ - 4}  \hfill.
\end{eqnarray}
According to Equation\,(\ref{eqn:utd}), the trap depth is monotonely
decreasing function of $\delta$ which is shown in
Fig.\,\ref{fig:trapdepth}. On the other hand, in order to reduce the
loss due to the atoms overflowing from the guide, the trap depth
should be larger than the mean atomic dynamic energy. This leads an
empirical condition \cite{Reichel02}
\begin{equation}
U_{i,j}>\eta k_B T, \label{eqn:trapcondition}
\end{equation}
with $\eta=5-7$ in order to make this loss term negligible. In order
to increase the trap depth, $l$ and $\delta$ should be reduced and
$I_0$ should be increased. Once another situation is added, another
limit will appear on selection of guide parameter which does not
exist in common static magnetic trap. According to
Section\,\ref{sec:1}, the Zeeman state before the atoms are trapped
is different from the adiabatic rf trapped state. When atoms are
entering the guide, there will be a probability of transferring
atoms from Zeeman state to adiabatic rf trapped state. According to
Equation (\ref{eqn:chi1}, \ref{eqn:chi2}, \ref{eqn:theta}), The key
parameter of transferring probability is $\delta /\Omega$, the
bigger $\delta /\Omega$ will increase the transferring probability.
The denominator, Rabi frequency $\Omega$, is different in different
spatial location because decided by amplitude of rf magnetic field
which is a quadrupole form and not homogenous. On the other hand,
the numerator $\delta$ is homogenous for a given guide. In both
cases there will appear a spatial distribution of transferring
probability. If atom cloud in $|F=2,m_{F}=-2\rangle$ Zeeman state is
put in the guide center (after performed optical pumping very fast)
and then adiabatic rf potential appears, the atoms near the guide
center will derive a high transferring probability and the farther
the atoms is from the guide center, the smaller transferring
probability is. As shown in Fig.\,\ref{fig:eff}, the transferring
probability of four-wire configuration is a little smaller than that
of three-wire configuration, and the smaller atom cloud will derive
the higher transferring probability. Also the bigger detuning
$\delta$ will cause the higher transferring probability. But
according to Fig.\,\ref{fig:trapdepth} the bigger detuning $\delta$
will cause the smaller trap depth, there is a contradiction when we
select the parameter $\delta$. We can not get the deeper trap depth
by increasing current amplitude or reducing spacing between wires
which will increase the Rabi frequency $\Omega$ and reduce the
transferring probability, so the simple idea for keeping trap depth
and increasing transferring probability is reducing the size of atom
cloud which can be realized by being compressed in a static magnetic
trap or an optical dipole trap and then pumping to
$|F=2,m_{F}=-2\rangle$ Zeeman state. For example, if the size of the
atom cloud $\sigma_x=\sigma_z=10\mu m$ and we want to obtain eighty
percent transferring probability, the detuning $\delta$ should be
greater than $2\pi\times7.6MHz$ and the trap depth is $117\mu K$
(four-wire) and $177\mu K$ (three-wire) along Z axis (the trap depth
along X axis is much bigger than it along Z axis). According to
Equation (\ref{eqn:trapcondition}), the temperature of atom cloud
should be smaller than $16\mu K$ and coincide with guide center
precisely.

\section{Design An Atom Chip with Adiabatic Radio-Frequency Guide}
\label{sec:3}

In common atom chip, almost all of parts is made from
nonferromagnetic material and can be considered as transparent ones
for magnetic field since they nearly do not change distribution of
dc magnetic field. In contrast to it, the metal accessories around
the rf atom chip must be considered because metal will change
distribution of rf electromagnetic field since metal is a kind of
good conductor. In order to create proper distribution of rf
magnetic field, we should design the chip configuration carefully.
Our atom chip is made by the technique developed by J. Reichel et al
\cite{Reichel01}, as shown in Fig.\,\ref{fig:section}(c), the fist
layer is made from silver and about $200nm$ in thickness; the second
layer is a kind of epoxy (Epotek 920)for fixing the silver layer and
keeping gold wires on the chip away from the silver layer.
Afterwards gold wires fabricated on an aluminium nitride chip base.
Below the base, there is an oxygen-free copper mount for fixing the
chip and elimination of heat. The epoxy and base is insulator and
not ferromagnetic material, so they nearly do not effect the
distribution of rf magnetic field. Whereas the silver layer and the
copper mount are both metal material, their effect on rf magnetic
field should be considered. In common chips, wire width $W=100\mu
m$, wire spacing $l=150\mu m$, wire thickness $d=8\mu m$ and
parameters about rf magnetic field, current amplitude $I_0=2A$,
detuning $\delta=2\pi\times7.6MHz$ and $B_0=0.055mT$ corresponding
to $\omega\approx2\pi\times10MHz$ which wavelength
$\lambda_{rf}=30m$ and skin depth is about $200\mu m$ for copper and
silver. The silver layer of 200nm in thickness is almost transparent
for the electromagnetic wave that we used although silver has very
good conductivity. On the other hand, size of the copper mount is
much bigger than the skin depth. After put onto a copper mount, rf
magnetic field created by the chip will change. It is nearly
impossible to calculate the distribution of the rf magnetic field by
analytic solution. FEM (Finite Element Method) is used for obtaining
numeric solution. As shown in Fig.\,\ref{fig:mounteffect}, position
of guide center goes far away from the chip surface when spacing
between copper mount and chip $d_{B}$ decreases. Afterwards the
adiabatic rf potential approaches the ideal configuration (without
width and copper mount) when $d_{B}\geq1900 \mu m$ and some results
are shown in Fig.\,\ref{fig:mounteffect} which show the effect of
mount. But this approximation is only true near the guide center and
great error will appear away from the guide center.

\section{Conclusion}
\label{sec:4}

We have shown our rf atom chip to be applicable to realize a guide
for atoms in strong-field seeking state. This technique may offer a
new way for manipulating atoms on an atom chip and potential
application in realizing atom interferometer and other fields in
integrated atom optics. The 2D character may be used to realize
coherent atom laser. If combined with surface induced evaporative
cooling technique \cite{Harber03}, an atom laser with continuous
output may be realized.

We acknowledge financial support from the State Key Basic Research
Program under Grant No. 2006CB921202, the National Natural Science
Foundation of China under Grant No. 10334050, 10474105 and Key
Oriental Project of Chinese Academy of Sciences under Grant No.
KGCX2-SW-100.

\newpage

\section*{List of Figure Captions}
Fig. 1. Section figure of the atom chip. Gold wires with rectangular
section (wire width w and height d) are fabricated on a chip and
one-wire configuration is showed in (a) where the wire on the chip
carries Rf current ($I_{s}=I_{0}sin(\omega t)$). In three-wire
configuration (b), $I_{L}=I_{R}=I_{0}sin(\omega t),
I_{M}=I_{0}sin(\omega t+\pi)$. In four-wire configuration
(c),$I_{a}=I_{c}=I_{0}sin(\omega t), I_{b}=I_{d}=I_{0}sin(\omega
t+\pi)$

Fig. 2. Contour figure of adiabatic rf potential created by a 3-wire
configuration where separation between two wires $l=150\mu m$,
current amplitude $I_{0}=2$A and current frequency detuing
$\delta=1$MHz.

Fig .3. Trap depth of trapped dressed state $|\chi_{1}\rangle$ in
three-wire and four wire configuration along X and Z axis where
$l=150\mu m$, current amplitude $I_{0}=2$A and unit of trap depth is
micro-Kelvin.

\begin{figure}
\centering
\includegraphics[width=8cm]{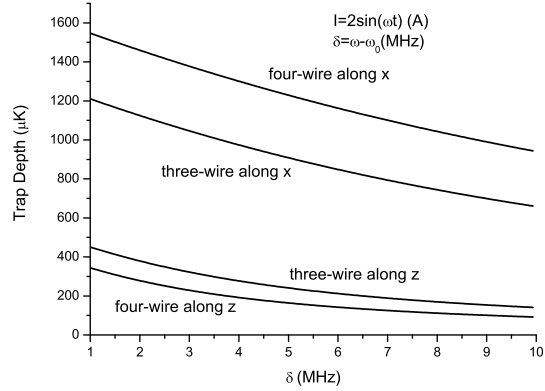}
\caption{Trap depth of trapped dressed state $|\chi_{1}\rangle$ in
three-wire and four wire configuration along X and Z axis where
$l=150\mu m$, current amplitude $I_{0}=2$A and unit of trap depth is
micro-Kelvin.} \label{fig:trapdepth}
\end{figure}

\begin{figure}
\centering
\includegraphics[width=8cm]{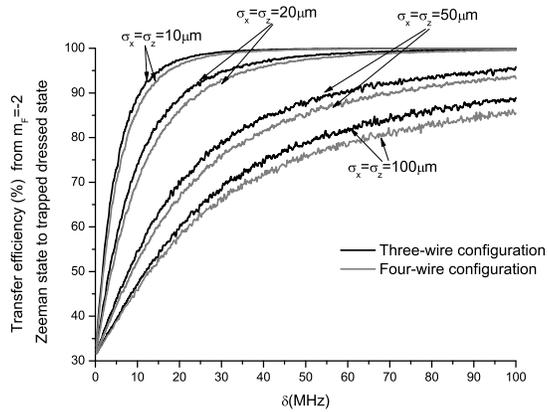}
\caption{Transfer efficiency (calculated by Monte Carlo Simulation)
from $|\varphi_{-2}\rangle$ ($|F=2,m_{F}=-2\rangle$-like state) to
all of the trapped dressed states($|\chi_1\rangle$ and
$|\chi_2\rangle$) in the three-wire and four-wire configuration and
its comparison among four kinds of atom cloud size (gaussian radius
$\sigma_{x}=\sigma_{z}=10\mu m, 20\mu m, 50\mu m,100\mu m$).
$\Omega$ is decided by rf current and configuration of wires on the
chip. Other parameters are identical to those in
Fig.\,\ref{fig:trapdepth}} \label{fig:eff}
\end{figure}

\begin{figure}
\centering
\includegraphics[width=8cm]{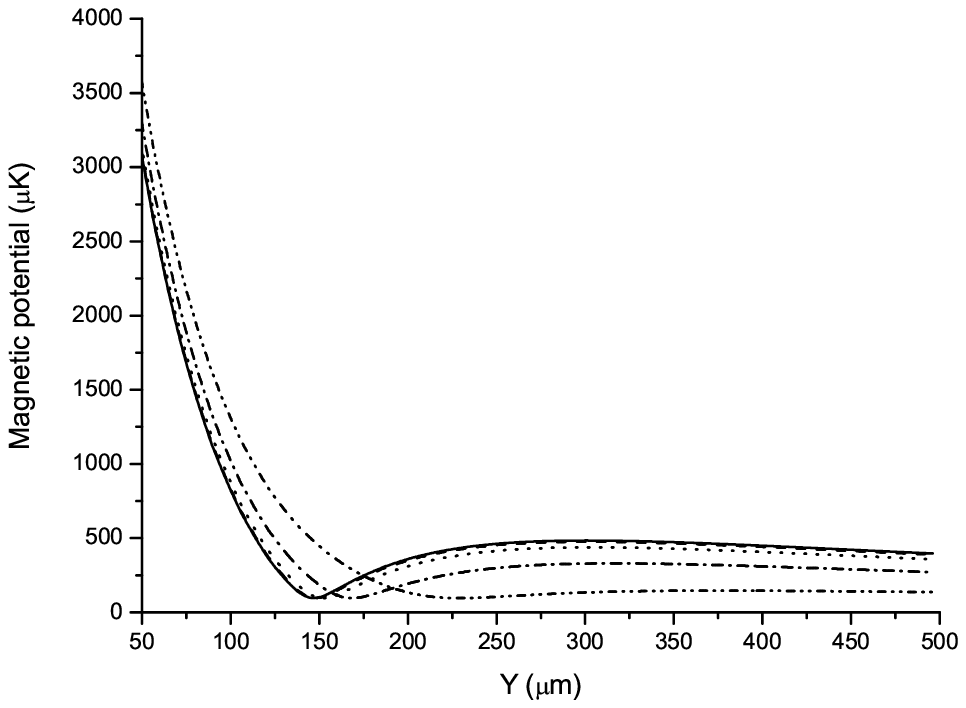}
\includegraphics[width=8cm]{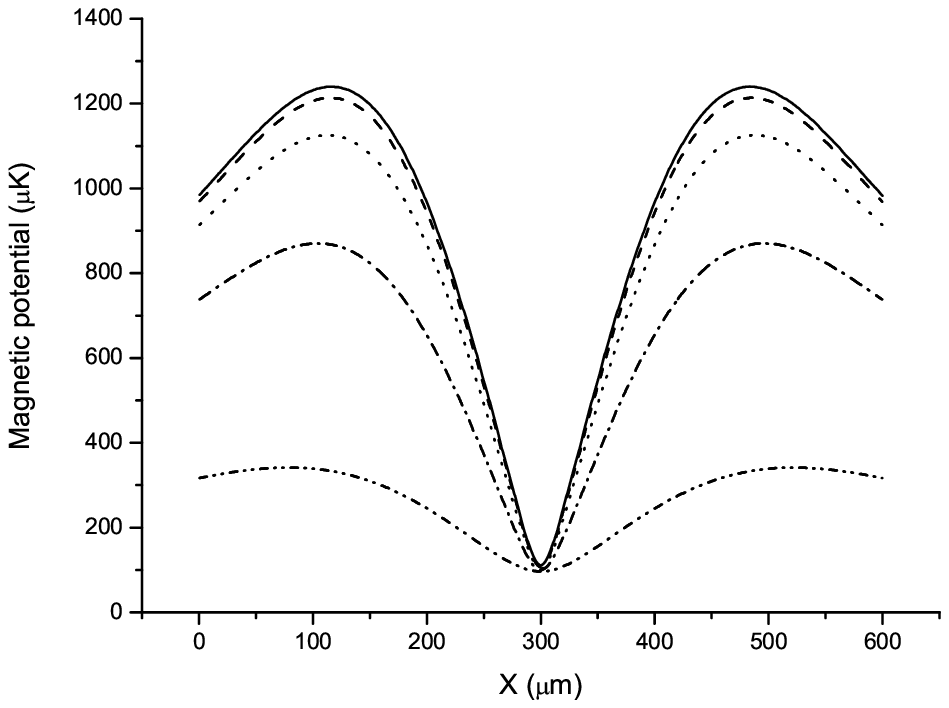}
\caption{Copper mount effect on magnetic potential of trapped
dressed state $|\chi_{1}\rangle$ in four wire configuration along Z
axis (a) and X axis (b), wire width $W=100\mu m$, wire spacing
$l=150\mu m$, wire thickness $d=8\mu m$, amplitude of rf Current
$I_0=2A$, rf field detuning $\delta=2\pi\times1MHz$, and distance
between chip and its copper mount is $100 \mu m$ (dash dot dot
plot), $500\mu m$ (dash dot plot), $1200\mu m$ (dot plot), $1800\mu
m$ (dash plot) and without mount (solid plot) respectively.}
\label{fig:mounteffect}
\end{figure}

Fig. 4. Transfer efficiency (calculated by Monte Carlo
Simulation)from $|\varphi_{-2}\rangle$($|F=2,m_{F}=-2\rangle$-like
state) to all of the trapped dressed states($|\chi_1\rangle$ and
$|\chi_2\rangle$) in the three-wire and four-wire configuration and
its comparison among four kinds of atom cloud size (gaussian radius
$\sigma_{x}=\sigma_{z}=10\mu m, 20\mu m, 50\mu m,100\mu m$).
$\Omega$ is decided by rf current and configuration of wires on the
chip. Other parameters are identical to those in
Fig.\,\ref{fig:trapdepth}

Fig. 5. Copper mount effect on magnetic potential of trapped dressed
state $|\chi_{1}\rangle$ in four wire configuration along Z axis (a)
and X axis (b), wire width $W=100\mu m$, wire spacing $l=150\mu m$,
wire thickness $d=8\mu m$, amplitude of rf Current $I_0=2A$, rf
field detuning $\delta=2\pi\times1MHz$, and distance between chip
and its copper mount is $100 \mu m$ (dash dot dot plot), $500\mu m$
(dash dot plot), $1200\mu m$ (dot plot), $1800\mu m$ (dash plot) and
without mount (solid plot) respectively.

\newpage

\end{document}